\documentstyle{article} 
\begin{document} 
\title{Quantum Mechanical  Embedding of Spinning Particle and Induced Spin-connection } 
\author{Naohisa OGAWA \thanks{E-mail: ogawa@particle.phys.hokudai.ac.jp}\\
Physics Department, Hokkaido University of Education, Sapporo 002  }
\maketitle 
\begin{abstract}
This paper introduces the way of the embedding of spinning particle  quantum mechanically for  non-relativistic case.  Schr\"odinger equation on its submanifold obtains the gauge field as spin connection, and it reduces to the ones obtained by Ohnuki and Kitakado when we consider $S^2$ in $R^3$. 
\end{abstract}
\section{Hypothesis and Schr\"odinger equation}
 Let us consider the motion of non-relativistic spinning particle on 
$M^{D-1}$ embedded in $ R^D$ quantum mechanically. There are two different ways to constrain the particle motion onto the submanifold. The first one is so called the confining potential approach \cite{Sch}, 
and the second one is the well known Dirac's procedure \cite{Dir}.  
In this paper we would like to work with the second procedure.
It should be stressed that it is straightforward to work in this scheme for spinless particle \cite{Dir},  but for spinning particle it is not known how to treat, and constrain the spin variables.  Therefore we propose somewhat intuitive quantization method by taking into account the essence of Dirac's scheme.  In non-relativistic mechanics the particle's spin is not related to the dynamics.  So we consider the embedding of dynamics separately to its spin.  This is the usual Dirac's treatment and giving the Hamiltonian as  Laplace-Beltrami (L.B.) operator plus quantum potential term.  But in this paper we do not concern with quantum potential term.  Then we obtain the Schr\"odinger equation but remaining the spin-index ``A" for wave function.
\begin{equation}
  i \hbar \frac{\partial \tilde{\Psi}^A(q,t)}{\partial t} = -\frac{\hbar^2}{2 m} K(\partial_q, q) ~ \tilde{\Psi}^A(q,t),
\end{equation}
where $K$ is the Laplace-Beltrami (L.B.) operator on $M^{D-1}$, and we write the local coordinates on $M^{D-1}$ as $q^{\mu}$ with $\{ \mu, \nu, \cdots = 1, \cdots D-1\}$.  The spin index  ``A" is the $SO(D)$ -spin index which is based on the rotational group in original space $R^D$, and its index is the same as the coordinate index: $x^A$  when we work with the adjoint representation which we use in the following.

 Now we consider the reduction of spin-index. In Dirac's treatment for second-class system, the quantization is done only for the independent variables, and there is no degree of freedom for normal direction to the hyper surface.  This relation should also be hold for spin variable, because the spin is  one representation of space rotation which is reduced to the smaller one by constraint.  Therefore the spin of wave function should reduce from $SO(D)$ to $SO(D-1)$ which is the rotation on tangent frame. This requirement is realized by relating the above wave function with the wave function $\Psi^a(q, t)$ with $SO(D-1)$ spin index  (or saying ``local Lorentz" indices) $\{a, b, \cdots = 1, \cdots D-1\}$ in real representation by the relation as
\begin{equation}
  \tilde{\Psi}^A(q,t ) = f^A_\mu(q) ~ h^\mu_a(q) ~\Psi^a(q, t) ,
\end{equation}
which is the key hypothesis in this paper.  
Here $f^A_\mu $ is the natural frame defined by
\begin{equation}
  f^A_\mu (q) \equiv  \frac{\partial x^A}{\partial q^\mu}.
\end{equation}
And $h^\mu_a(q)$ is the vielbein defined on  $M^{D-1}$  by
\begin{equation}
f^A_\mu f_{A \nu} \equiv  g_{\mu \nu} = \delta_{a b} ~h^a_
 \mu~ h^b_\nu,
\end{equation}
\begin{equation}
   h^a_\mu h^\mu_b = \delta^a_b, ~~ h^a_\mu h^\nu_a = \delta^\nu_\mu.
\end{equation}
where $g_{\mu \nu}$ is the induced metric on $M^{D-1}$ . 
The Indices $\{\mu, \nu, \cdots \}$ are raised and lowered by induced metric  $g_{\mu \nu}, ~ g^{\mu \nu}$,  ``local Lorentz" indices $\{ a, b, \cdots \}$ are raised and lowered by $\delta_{a b}, ~ \delta^{a b}$ , and Euclidean indices $\{A, B, \cdots\}$ are also raised and lowered by 
$\delta_{A B}, ~ \delta^{A B}$ .  

The essence of this hypothesis is that the vector index ``A"  has  only the tangential component, and normal component is vanishing.  This vanishing of normal component insures the vanishing of space rotation on planes constructed between normal and tangent axes at each point on $M^{D-1}$ as we discussed earlier.
In relativistic theory of the same problem,  since there is classical object corresponding to spin: supersymmetry, we can perform the Dirac's procedure, and by using the current conservation we can prove the above hypothesis \cite{rel}.  
But for our non-relativistic case we can not prove it, and we should remark that above hypothesis means one assumption essentially, that is,  the normal component of wave function is vanishing by constraint.  The remained tangential components of wave function is related to the local Lorentz index of the new wave function by the hypothesis.  
But in other words, it is the definition of the new wave function  
$\Psi^a(q, t)$, which is introduced just to make clear the assumption. 
 We will explain this point more explicitly.   
By introducing the curvilinear coordinates in $R^N$, we consider the coordinate transformation as
$$     \{x^A\}  \to  \{q^{\bot}, q^\mu \} .  $$
And we choose the coordinate that $M^{D-1}$ is defined by 
$q^{\bot} = c$, where the $q^{\bot} $ axis is normal to that manifold.
Then the assumption discussed above is written as
$$  \frac{\partial q^{\bot} }{\partial x^A} ~\tilde{\Psi}^A(q^\nu)  = 0.$$
This condition requires the form for wave function by using some wave function $\phi^\mu$ as
$$     \tilde{\Psi}^A  =  f^A_\mu ~\phi^\mu.  $$
We can identify $\phi^\mu$  as the vector field on $M^{D-1}$ which can always be rewritten to the field with local Lorentz index by vielbein as
$$  \Psi^a = h^a_\mu \phi^\mu, ~~~~~\phi^\mu = h^\mu_a \Psi^a.$$
This is the definition of $\Psi^a$ and we obtain our hypothesis from one assumption. The reason why we utilize $\Psi^a$ but not $\phi^\mu$ as the independent wave function is just the favor, and we can use latter one also as the equivalent object.   The difference is only the way of representation.
By putting (2) into (1), and multiplying $f^A_\nu h^{\nu b}$ from left hand sides, and also by using the notation and relation
\begin{equation}
   \frac{\partial}{\partial x^A} \equiv  
   f_A^\mu ~ \frac{\partial}{\partial q^\mu},~~~ 
\frac{\partial^2}{\partial \vec{x}^{\, 2}} = K(\partial_q, q),
\end{equation}
we obtain the results,
\begin{equation}
 i \hbar \frac{\partial \Psi^b}{\partial t} = -\frac{\hbar^2}{2 m}  ~[~\sum_A (\delta^b_d \frac {\partial}{\partial x^A} + \omega^b_{~d, \mu} f^\mu_A)  (\delta^d_c \frac {\partial}{\partial x^A} + \omega^d_{~c, \nu} f^\nu_A) -  (H^2)^b_{~c} ~]~ \Psi^c,
\end{equation}
where $\omega^b_{~d, \mu} $ is the spin connection which is related to the ``local Lorentz" symmetry  hidden in (4) , that is, the local $SO(D-1)$ rotation on each tangent spaces attached on $M^{D-1}$, and its form is the same as the ones defined by usual vielbein hypothesis as
\begin{equation}
  {\cal D}_\mu h_{\nu a} \equiv \partial_\mu h_{\nu a} + \omega_{a b,\mu}h^b_\nu - \Gamma^{\lambda}_{\nu \mu} h_{a \lambda} = 0,
\end{equation}
or more explicitly,
\begin{equation}
\omega^a_{~b,\mu} = h^{\nu a} \partial_\mu h_{\nu b} - \Gamma^{\lambda}_{\nu \mu} h^{\nu a} h_{\lambda b}.
\end{equation}
Notice that once we obtained the Schr\"odinger equation (7), we can  change to the general representation by using the spin matrix ${\bf S}$ except the last term. Then the gauge term takes the form:
$$ f^\nu_A ~\omega^{dc}_\nu ~{\bf S}_{dc}~ \Psi(q).$$
 The last ``quantum potential" term has the sense
\begin{equation}
      (H^2)^b_{~c} \equiv h^{\mu b} H_{\mu \nu} H^{\nu \lambda} 
h_{\lambda c},
\end{equation}
with $H_{\mu \nu}$ as the second fundamental tensor :  extrinsic curvature defined by
\begin{equation}
     H_{\mu \nu} = \sum_A ~\partial_\nu f^A_\mu  ~n^A.
\end{equation}
Here $ n^A$ is the normal unit vector to $M^{D-1}$. 
The set $ f^A_i = \{n^A, f^A_\mu \}$  ($i = 0, 1, \cdots D-1 $) forms the complete set on $M^{D-1}$ in $R^D$, that is,
\begin{equation} 
 g_{i j} = \sum_A ~f^A_i~ f^A_j = \left( 
\begin{array}{c|ccc}
1  &   & 0 & \\
\hline
   &   &   &  \\
0 &   & g_{\mu \nu} & \\
   &   &   & 
\end{array}
 \right), 
\end{equation}
\begin{equation} 
  g^{ij} ~f^A_i ~ f^B_j ~ = ~ \delta^{A B}. 
\end{equation}
The latter relation was utilized to derive (7). 
Note that this kind of $\hbar^2$-quantum potential usually appears in the precise treatment of embedding, and we do not devote to this point in this paper \cite{Sch},\cite{Dir},\cite{Oga}. Instead we would like to take attention here especially to the induced  spin connection.
\section{Form of Connection}
   Let us estimate the explicit form of our connection in the case of
  $S^2$ in $R^3$.
We use the metric
\begin{equation}
ds^2 = g_{\mu \nu} dq^\mu dq^\nu = R^2 d\theta^2 + R^2 \sin^2 \theta ~d\phi^2, ~~ q^1 = \theta,~q^2 = \phi.
\end{equation}
Then we can take the following form for $h^a_{\mu}$.
\begin{eqnarray}
  h^1_1 &=& R \cos \delta, ~~~~~h^2_1 = -R \sin \delta,\\
  h^1_2 &=& -R \sin \theta  \sin \delta, ~~
  h^2_2 = -R \sin \theta  \cos \delta,
\end{eqnarray}
where $\delta$ is the any function of $\theta$ and $\phi$ as the gauge choice.  We calculate only $\omega^{21}_\mu$ in the following, since its spin index is antisymmetric. The result is that
\begin{equation}
  \omega^{21}_1 = \partial_{\theta} \delta,~~~
 \omega^{21}_2 = \partial_{\phi} \delta - \cos \theta.
\end{equation}
Note that we can not vanish all the connection by choosing the gauge: $\delta (\theta, \phi)$.   Using the coordinate condition as 
\begin{equation}
x^1 = R~ \sin \theta~\cos \phi, ~~x^2 = R~ \sin \theta~\sin \phi,
~~x^3 = R~\cos \theta,
\end{equation}
$\vec{f}^\mu = f^\mu_A$ takes the form.
\begin{eqnarray}
 \vec{f}^{~1} &=& R^{-1} ( \cos \theta~\cos \phi,~\cos \theta~\sin \phi,~-\sin \theta), \\
 \vec{f}^{~2}  &=& R^{-1} ( -\frac{\sin \phi}{\sin \theta} ,~~\frac{\cos \phi}{\sin \theta},~~0).
\end{eqnarray}
Then the induced gauge field: $ \vec{A} \equiv \omega^{21}_\mu \vec{f}^\mu$ takes the form:
\begin{eqnarray}
\vec{A} &=& R^{-1} [  \cos \theta ~\cos \phi ~\partial_\theta \delta + 
\frac{\sin \phi}{\sin \theta} (\cos \theta - \partial_\phi \delta), \nonumber \\
&& \cos \theta \sin \phi ~\partial_\theta \delta - \frac{\cos \phi}{\sin \theta} (\cos \theta - \partial_\phi \delta),
~~-\sin \theta ~\partial_\theta \delta].
\end{eqnarray}
One simple gauge choice is to take $\delta = \phi$.  Then our gauge field takes the following monopole-like configuration.
\begin{equation}
\vec{A} = [\frac{-y}{R(R + z)},~~\frac{x}{R(R+z)},~~0].
\end{equation}
This structure of gauge field may be universal in the kind of embedding like $S^{D-1}$ in $R^D$.  Really this kind of gauge field with embedded particle on $S^{D-1}$ was firstly obtained by Ohnuki and Kitakado in different way \cite{Ohn}. 
From the above consideration, we conclude that their gauge field will be
essentially the same as the spin connection for the spinning particle embedded on $S^{D-1}$ .

\section{Discussion}
Lastly we would like to give some comments to our starting quantization scheme. There are three kinds of embedding procedure. One is due to the Dirac's one \cite{Dir} (reduced phase space method) where all the normal degree of freedom is frozen in each operator variables.  Another one is due to  the Schr\"odinger equation with confining potential method (abbreviated as CP approach hereafter)\cite{Sch}, where all the normal degree of freedom is alive as well as tangential one, and the connection related to the rotation of normal basis appears.  The last one is due to the group theoretical approach \cite{Ohn}, where the geometrical connection related to the rotation of tangential basis appears.  
In the case of spin less particle, we can estimate first two approaches and the difference appears quantum mechanically at the quantum mechanical potential \cite{Oga}, and the disappearance and appearance of connection.  Instead in the case of spinning particle, Dirac's scheme can not be used  directly since there is no classical object like particle's spin. In CP approach there is no interesting effect for spin. This is because, in that scheme there remains normal (to submanifold)  degree of freedom, and spin can still take any direction in $R^D$ (spin is still in the $SO(D)$ representation).  This means that the embedding of spin can not be done in CP approach, and the connection obtained in this article can not be obtained in that scheme.  In this sense, it is interesting  to construct the Dirac's embedding scheme  for spinning particle, where spin degree is also reduced onto the embedding manifold, and it is expected that its result have the same form as the one in this article.     Our approach in this article and group theoretical approach is much different, but their physical contents may be equivalent at least when we consider $S^{D-1}$ in $R^D$.  Because both have the connection typed $SO(D-1)$ related to the rotation of tangential basis.  The connection obtained here and its formalism is one physical interpretation of ohnuki-kitakado's connection.
\par
\

\noindent{\em Acknowledgement.}
The author especially wish to thank Prof.K.Fujii for his useful discussion on Ohnuki-Kitakado's connection, and also thank Prof.T.Okazaki for usual encouragement to him.


\begin{thebibliography}{10}
\bibitem{Sch} R.C.T.da Costa, Phys.Rev. {\bf 23} (1981), p1982;  J.Tolar, 1988 Lecture Notes in Physics {\bf 313}, ed. H.D.Doever, J.D.Henning and T.D.Raev \\ (Springer-Verlag, Berin, Heiderberg), p.268;  M.Ikegami and Y.Nagaoka, Prog.Theor.Phys. Suppl.  {\bf 106} (1991),  p235;   S.Takagi and T.Tanzawa, Prog.Theor.Phys. {\bf 87} (1992), p561; K.Fujii and N.Ogawa, Prog.Theor.Phys. {\bf 89} (1993), p575;  P.Maraner, J.Phys. {\bf A28}, 2939, 1995; P.Maraner Ann.Phys. {\bf 246}, 325, 1996;
 K.Fujii and S.Uchiyama Prog.Theor.Phys.{\bf 95} (1996), p461.
\bibitem{Dir}  N.Ogawa, K.Fujii, and A.Kobushkin, Prog.Theor.Phys. {\bf 83} (1990), p894; N.Ogawa, K.Fujii, N.Chepilko and A.Kobushkin, Prog.Theor.Phys. {\bf 85} (1991), p1189.
\bibitem{Oga}  N.Ogawa, Prog.Theor.Phys. {\bf 87}(1992), p513.
\bibitem{Ohn}  Y.Ohnuki and S.Kitakado, J.Math.Phys. {\bf 34} (1993), p2827; \\ K.Fujii, S.Kitakado and Y.Ohnuki, Mod. Phys. Lett. {\bf A10} (1995), p867;  see also for general view point, D.McMullan and I.Tsutsui, Ann. of Phys.{\bf 237} (1995), p269; Phys.Lett.{\bf 320B}(1994), p287.
\bibitem{rel} N.Ogawa, Embedding of Relativistic Particles and Spinor Natural-Frame,  hep-th/9703181  , Mar. 1997.
\end{thebibliography}
\end{document}